\documentclass[aps,onecolumn,10pt,superscriptaddress,floatfix,nofootinbib]{revtex4}
\usepackage{subfigure}
\usepackage{epsfig}
\usepackage{epstopdf}
\usepackage{color}
\usepackage{graphicx}
\usepackage{longtable}
\usepackage{CJK}
\usepackage{amsmath,bbold}
\usepackage{float}
\usepackage{physics}
\usepackage{comment}

\usepackage{hyperref}
\hypersetup{colorlinks=true,linkcolor=red, citecolor=blue,  urlcolor=blue}
\makeatletter
\def\@yfootnote{\stepcounter\@mpfn\protected@xdef\@thefnmark{\thempfn}\@footnotemark}
\makeatother

\usepackage{listings}
\usepackage{xcolor}

\definecolor{codegreen}{rgb}{0,0.6,0}
\definecolor{codegray}{rgb}{0.5,0.5,0.5}
\definecolor{codepurple}{rgb}{0.58,0,0.82}
\definecolor{backcolour}{rgb}{0.95,0.95,0.92}

\lstdefinestyle{mystyle}{
    backgroundcolor=\color{backcolour},
    commentstyle=\color{codegreen},
    keywordstyle=\color{magenta},
    numberstyle=\tiny\color{codegray},
    stringstyle=\color{codepurple},
    basicstyle=\ttfamily\footnotesize,
    breakatwhitespace=false,
    breaklines=true,
    captionpos=b,
    keepspaces=true,
    numbers=left,
    numbersep=5pt,
    showspaces=false,
    showstringspaces=false,
    showtabs=false,
    tabsize=2
}

\begin{document}

\title{\texttt{u3lib}: An AI-Ready Computational Framework for U(3) Symmetry Algebra,\\ Wigner Coupling, and Racah Recoupling Coefficients}

\author{Daniel Langr}
\affiliation{Department of Computer Systems, Faculty of Information Technology, Czech Technical University in Prague, Prague, Czech Republic}

\author{Phong Dang}
\affiliation{Department of Physics and Astronomy, Louisiana State University, Baton Rouge, Louisiana, USA}

\author{Tomas Dytrych}
\affiliation{Nuclear Physics Institute, Czech Academy of Sciences, \v{R}e\v{z}, Czech Republic}
\affiliation{Department of Physics and Astronomy, Louisiana State University, Baton Rouge, Louisiana, USA}

\author{Michela Negro}
\affiliation{Department of Physics and Astronomy, Louisiana State University, Baton Rouge, Louisiana, USA}

\begin{abstract}
\texttt{u3lib} is a C++ library for U(3)/SU(3) representation-theory
calculations in nuclear structure and related applications. It evaluates
Wigner coupling and Racah recoupling coefficients using the procedure proposed in 2024
by Dang \textit{et al.}, which avoids the binomial coefficients and alternating
sums used in earlier formulations. The library also provides irreducible
representations (irreps), dimensions, Casimir invariants, conjugation, and
tensor-product decompositions with outer multiplicities. It supports the
canonical Gelfand-Tsetlin basis and the physical
$\mathrm{SU(3)}\supset\mathrm{SO(3)}$ basis, including inner multiplicities
and the transformation between the two bases. Generator routines return
matrix elements in the canonical basis and reduced matrix elements in the
physical basis. Python bindings mirror the C++ API. Domain skills supplied
with the repository allow an AI assistant with code-execution access to call
these routines from natural-language requests.

\end{abstract}


\maketitle

\section*{PROGRAM SUMMARY}
\noindent
\textit{Program title}: \texttt{u3lib}

\noindent
\textit{Licensing provisions}: MIT License.

\noindent
\textit{Code repository}: \url{https://github.com/Deep-SANE/u3lib}

\noindent
\textit{Programming language}: C++ (C++20), with Python bindings via pybind11.

\noindent
\textit{External libraries}: \texttt{Eigen3} for QR-based null space solver; \texttt{Boost} for hash tables;
\texttt{OpenMP} for multi-threading; \texttt{fmt} for formatting output; \texttt{libwignernj} for $\rm SU(2)$ coupling
and recoupling coefficients; and \texttt{KraNS} for iterative sparse null-space solvers on large matrices
(with an automatic dense-QR fallback for small or block-heavy cases).
The Python extension additionally uses \texttt{pybind11}.

\noindent
\textit{Nature of problem}: Fast, accurate, and reproducible evaluation of 
Wigner reduced coupling coefficients for the two group-subgroup chains
$\mathrm{SU(3)}\supset\mathrm{SO(3)}$ and
$\mathrm{U(3)}\supset\mathrm{U(2)}$, Racah
recoupling coefficients (6-$\rm U(3)$ and 9-$\rm U(3)$ coefficients), and
the associated algebraic quantities (irrep dimensions, Casimirs, multiplicities, 
basis enumeration, generator matrix elements, and basis transformations) needed by 
symmetry-adapted calculations for nuclear structure and similar problems.

\noindent
\textit{Solution method}: The $\mathrm{U(3)}\supset\mathrm{U(2)}$ reduced coupling and 
recoupling coefficients are built from the highest-weight $\mathrm{U(3)}\supset\mathrm{U(2)} \supset\mathrm{U(1)}$ 
full coupling coefficients, obtained as the orthonormal null space of the canonical raising generators, 
and propagated by the corresponding lowering generators \cite{Dang2024EPJP1}; 
the physical $\mathrm{SU(3)}\supset\mathrm{SO(3)}$ coefficients follow from the physical-to-canonical
basis transformation, also obtained as the orthonormal null space of the physical raising generator $L_+$ \cite{Dang2024EPJP1}.

\section{Introduction}
\label{sec:intro}

Algorithms for computing coupling and recoupling coefficients for the unitary group
$\rm U(3)$ and its special unitary subgroup $\rm SU(3)$ date back to two papers published in 1973 by Draayer and Akiyama,
which introduced a general algorithm \cite{Draayer1973JMP} and an
associated Fortran program for their evaluation \cite{Akiyama1973CPC}.
These coefficients underpin nuclear structure calculations using Elliott's
$\rm SU(3)$ model \cite{Elliott1958I,Elliott1958II} and other symmetry-adapted
shell-model approaches \cite{Dytrych2013PRL,Dytrych2020PRL,McCoy2020PRL}, where large numbers of coefficients make speed and accuracy essential.
The same symmetry is used in nuclear-structure frameworks for
medium-mass and heavy deformed nuclei, including the pseudo-$\rm SU(3)$
\cite{Kota2020,Cseh2020PRC,Gade2025NatPhys}, proxy-$\rm SU(3)$ \cite{Bonatsos2023Symmetry,Cseh2020PRC},
and quasi-$\rm SU(3)$ \cite{Zuker1995PRC,Gade2025NatPhys} schemes, as well as the
multiconfigurational dynamical symmetry (MUSY) \cite{Cseh2021PRC}. Over the following five decades, the Draayer
and Akiyama library was updated to improve numerical accuracy
\cite{Bahri2004CPC} and implemented in modern programming languages
\cite{Dytrych2021CPC,Herko2026EPJA}. These implementations retained the
underlying algorithm based on hypergeometric functions with large factorials
and alternating sums. Dang \textit{et al.} introduced a different approach that does 
not involve these instability-prone functions
by constructing the coefficients from the action of the group's raising generators
using numerical null-space solvers \cite{Dang2024EPJP1,Alex2011JMP}.
For the large sparse matrices arising in this approach,
a dedicated iterative null-space solver, now known as
\texttt{KraNS}, was developed \cite{Krastenov2026MSc}.

The same algebraic tools also arise outside nuclear physics. In particle
physics, flavor $\rm SU(3)$ underlies the eightfold-way classification of
hadrons and the quark model \cite{GellMann1962PR,Neeman1961NP}. In quantum
information, Schur transforms are constructed from successive Clebsch--Gordan
transformations for $d$-level systems; for $d=3$, this provides a potential
application of U(3) coupling data to qutrit basis transformations
\cite{Bacon2006PRL}. In condensed-matter physics, symmetry-adapted numerical
renormalization-group calculations of the SU(3) Anderson impurity model require
tensor-product decompositions, coupling coefficients, and
outer-multiplicity labels \cite{Moca2012PRB}. In quantum optics, SU(3)
representation theory organizes multiphoton states and interference in
three-channel passive interferometers \cite{Tan2013PRL}. In quantum simulations
of $\rm SU(3)$ lattice gauge theory, basis states are organized by $\rm SU(3)$
irreps and Gelfand--Tsetlin patterns, and the gauge-invariant singlet state at
each vertex is assembled from coupling coefficients \cite{Draper2026}. These
examples suggest
uses for the computational framework presented here as a source of
representation-theoretic building blocks, with application-specific modeling and
any basis conversion supplied by the user.

In this paper, we present \texttt{u3lib}, a library for calculations with the
$\rm U(3)$ symmetry algebra, including coupling and recoupling coefficients.
The library is part of
\texttt{DeepSANE}---\textbf{Deep}-learning and \textbf{S}ymmetry-adapted \textbf{A}b-initio 
\textbf{N}uclear modeling: an AI-ready \textbf{E}cosystem\footnote{See the \texttt{DeepSANE} GitHub organization: 
\url{https://github.com/Deep-SANE}}. 
Alongside the coefficient routines, \texttt{u3lib} provides C++ objects for irreps
and their Casimir invariants, the canonical
Gelfand-Tsetlin $\rm U(3) \supset U(2) \supset U(1)$ and the physical $\rm SU(3) \supset SO(3)$ bases, 
the generator matrix elements, the tensor-product decomposition, and the physical-to-canonical transformation. 
The C++ API is header-only; Python bindings mirror it one-to-one.

Large language models can help run scientific calculations from
natural-language requests. To support their use with \texttt{u3lib}, we provide
an agent-skill catalog that guides AI assistants through the library's
capabilities. To use these routines correctly, an agent must follow group-theoretical notation,
phase conventions, memory-lifetime rules, and thread-safety requirements.
The repository-local skills make these requirements explicit alongside the
source code and provide procedures for constructing and checking calculations.
This approach follows the interest in reusable domain-specific skills for
computational research seen in initiatives such as the U.S.
Department of Energy (DOE) Genesis Mission and its scientific agent skill 
catalog\footnote{See the DOE Genesis Mission skill catalog: \url{https://gitlab.osti.gov/genesis/genesis-skills}.}.
We also evaluate whether this guidance improves task completion and
reduces the resources needed to integrate the library, particularly as
implementation complexity increases. 

The remainder of this paper documents the repository structure
(Sec.~\ref{sec:structure}), the classes and interface (Sec.~\ref{sec:classes}),
the Python bindings (Sec.~\ref{sec:python}), the AI-ready interface and its
benchmark evaluation (Sec.~\ref{sec:agents}), the installation (Sec.~\ref{sec:build})
and a short summary (Sec.~\ref{sec:conclusion}).

\section{Library structure}
\label{sec:structure}

\texttt{u3lib} is a header-only C++ library with build targets for examples, tests, and Python bindings. The top-level
directories are:
\begin{itemize}
    \item \verb|include/|: the library itself (all algorithms live here; user
    programs include these headers directly):
    \begin{itemize}
        \item \verb|U3irrep.h|: the irrep classes \verb|su3::irrep|, \verb|u3::irrep|;
        \item \verb|U3gelfand.h|, \verb|internal/gelfand/|: the Gelfand state class \verb|u3::gelfand|;
        \item \verb|U3basis.h|: basis enumeration (\verb|gelfandBasis|, \verb|physicalBasis|), the inner multiplicity \verb|Kappa|, the \verb|u3::physicalvector| class, and the generator-action predicates;
        \item \verb|U3decomposition.h|: the outer multiplicity \verb|Rho| and the tensor-product \verb|decomposition|;
        \item \verb|U3U2xU3U2.h|, \verb|U3SO3xU3SO3.h|: the canonical and physical coupling coefficients and their look-up tables;
        \item \verb|U3_U6Z6Coeffs.h|, \verb|U3_U9Coeffs.h|: the 6-$\rm U(3)$ and 9-$\rm U(3)$ recoupling coefficients and their look-up tables;
        \item \verb|U3SO3_to_U3U2U1.h|: the physical-to-canonical transformation;
        \item \verb|U3generatorsRME.h|: the physical-generator reduced matrix elements \verb|RME_L|, \verb|RME_Q|;
        \item \verb|Definitions.h|: the structured coefficient containers;
        \item \verb|u3lib.h|: the umbrella header and the \verb|initialize|/\verb|finalize| lifecycle.
    \end{itemize}
    \item \verb|src/|: example driver programs (one \verb|.cpp| per task) that call the headers and print results; compiled to \verb|bin/|.
    \item \verb|python/|: the pybind11 extension (\verb|pyu3lib.cpp| $\to$ \verb|u3lib.so|, built with \verb|make python|) and a tutorial Jupyter notebook (\verb|tutorial.ipynb|).
    \item \verb|agents/|: the AI-agent skill catalog: an index (\verb|AGENTS.md|), shared runtime instructions, and one skill per capability.
    \item \verb|test/|: the regression suite (\verb|test_*.cpp|, the \verb|*_cases.h| case sets, and \verb|reference/| documented values), run by \verb|make test|.
    \item \verb|doc/|: this manual (LaTeX source).
    \item \verb|bin/|: build output (not source).
\end{itemize}

\section{Classes and library interface}
\label{sec:classes}

This section describes the public classes, methods, and interfaces for accessing
coefficients. Types belong to the \verb|u3| namespace, with SU(3) base types in
\verb|su3|. The algebraic types can be used without initialization. Coefficient
databases require initialization as described below.
Table~\ref{tab:main-funcs} lists the main entry points.

\begin{table*}[tp]
    \centering
    \begin{tabular}{p{3.2cm}p{5.8cm}p{6.2cm}}
    \hline\hline
         Header & Entry point & Task \\
        \hline
         \verb|U3irrep.h| & \verb|dim|, \verb|C1/C2/C3|, \verb|conjugate| & irrep dimension, Casimirs, conjugation \\
         \verb|U3basis.h| & \verb|Kappa|, \verb|gelfandBasis|, \verb|physicalBasis| & inner multiplicity; basis enumeration \\
         \verb|U3decomposition.h| & \verb|Rho|, \verb|decomposition| & outer multiplicity; irrep product  \\
         \verb|U3U2xU3U2.h| & \verb|wigU3U2db| & $\mathrm{U(3)}{\supset}\mathrm{U(2)}$ reduced coupling coefficients  \\
         \verb|U3SO3xU3SO3.h| & \verb|wigU3SO3db| & $\mathrm{SU(3)}{\supset}\mathrm{SO(3)}$ reduced coupling coefficients  \\
         \verb|U3_U6Z6Coeffs.h| & \verb|u3u6db|, \verb|u3z6db| & 6-$\rm U(3)$ recoupling coefficients (three irreps)  \\
         \verb|U3_U9Coeffs.h| & \verb|u3u9db| & 9-$\rm U(3)$ recoupling coefficients (four irreps)  \\
         \verb|U3SO3_to_U3U2U1.h| & \verb|phys_to_cano| & physical-to-canonical transformation \\
         \verb|U3generatorsRME.h| & \verb|RME_L|, \verb|RME_Q| & reduced matrix elements of $\mathbf{L}$, $\mathbf{Q}$ \\
    \hline\hline
    \end{tabular}
    \caption{Main entry points of \texttt{u3lib}.}
    \label{tab:main-funcs}
\end{table*}

\subsection{Lifecycle}
In \verb|u3lib.h|:
\begin{itemize}
    \item \verb|initialize()| / \verb|initialize(&argc,&argv)|: set up the caches of the $\rm SU(2)$ coefficients from \texttt{libwignernj}, and the U(3) coefficient databases;
    \item \verb|finalize()|: destroy the databases;
    \item \verb|initialize_thread()| / \verb|finalize_thread()|: per-thread setup for multithreaded use;
    \item \verb|clear()|: clear cached coefficients while keeping the databases available;
    \item \verb|is_initialized()|: check whether the library is initialized.
\end{itemize}

Note on shared-backend lifecycle: if \texttt{u3lib} runs in the same process as another library that shares the \texttt{libwignernj} backend, initialize all of them at startup and finalize only after all coefficient work is complete. Calling \verb|finalize()| while another user of the backend is still active destroys the shared state and causes runtime errors.

\subsection{Irreps: \texttt{su3::irrep} and \texttt{u3::irrep}}
(\verb|U3irrep.h|.) \verb|su3::irrep| stores an SU(3) irrep; \verb|u3::irrep|
derives from it and adds the third label $n_{33}$ (fixing the total quanta).

\textbf{Constructors:}
\begin{itemize}
    \item \verb|su3::irrep()|: default (empty) SU(3) irrep.
    \item \verb|su3::irrep(lambda,mu)|: SU(3) irrep from its labels.
    \item \verb|u3::irrep()|: default (empty) U(3) irrep.
    \item \verb|u3::irrep(n13,n23,n33)|: U(3) irrep from its three labels.
    \item \verb|u3::irrep(lambda,mu)|: normalized representative $[\lambda{+}\mu,\mu,0]$.
    \item \verb|u3::irrep(su3::irrep, N)|: fix the total quanta $N$.
\end{itemize}

\textbf{Label getters:}
\begin{itemize}
    \item \verb|lam()| / \verb|lambda()|: the SU(3) label $\lambda$.
    \item \verb|mu()|: the SU(3) label $\mu$.
    \item \verb|n13()|, \verb|n23()|, \verb|n33()|: the three U(3) labels (only for \verb|u3::irrep|).
\end{itemize}

\textbf{Dimension and invariants:}
\begin{itemize}
    \item \verb|dim()|: the dimension (both classes).
    \item \verb|C2()|: the quadratic SU(3) Casimir (both classes, shift-invariant).
    \item \verb|C3()|: the cubic SU(3) Casimir (both classes, shift-invariant).
    \item \verb|C1()|: the first-order Casimir $=n_{13}{+}n_{23}{+}n_{33}$ (total quanta), on \verb|u3::irrep| only; \verb|su3::irrep| has no \verb|C1()|.
    \item \verb|quanta()|: synonym of \verb|C1()|.
\end{itemize}

\textbf{Other:}
\begin{itemize}
    \item \verb|zero()|: test for the trivial irrep.
    \item \verb|conjugate()|: the conjugate irrep $(\mu,\lambda)$.
    \item \verb|conjugate_phase()|: the associated conjugation phase.
    \item \verb|normalize()|: set $n_{33}=0$ for \verb|u3::irrep|.
    \item comparison operators and stream output.
    \item \verb|writeBinary| / \verb|readBinary|: serialization for binary files.
\end{itemize}

\subsection{Gelfand states: \texttt{u3::gelfand}}
(\verb|U3gelfand.h|.) The canonical basis state $\ket{n_{13},n_{23},n_{33};\,n_{12},n_{22};\,n_{11}}$.

\textbf{Constructor:}
\begin{itemize}
    \item \verb|u3::gelfand(n13,n23,n33,n12,n22,n11)|: build a Gelfand state from its labels.
\end{itemize}

\textbf{Label getters and validity check:}
\begin{itemize}
    \item \verb|n13()| \dots\ \verb|n11()|: the six Gelfand-pattern labels.
    \item \verb|check()|: test the betweenness condition.
    \item \verb|normalize()|: shift all quantum numbers by $n_{33}$.
\end{itemize}

\textbf{Subgroup labels:}
\begin{itemize}
    \item \verb|lam()| / \verb|lambda()|: the SU(3) label $\lambda$.
    \item \verb|mu()|: the SU(3) label $\mu$.
    \item \verb|epsilon()|: the U(1) weight.
    \item \verb|Lambdax2()|: $2\times$ the embedded SU(2) angular momentum.
    \item \verb|MLambdax2()|: $2\times$ its projection.
\end{itemize}

\textbf{Conjugation:}
\begin{itemize}
    \item \verb|conjugate()|: the conjugate Gelfand state.
    \item \verb|conjugate_phase()|: the associated conjugation phase.
\end{itemize}

\textbf{Generator matrix elements:} Each returns the matrix element of a U(3)
canonical generator $E_{ij}$ acting on a Gelfand state; a \verb|_12|/\verb|_22| suffix selects the
target branch (either on $n_{12}$ or $n_{22}$, see \cite{Dang2024EPJP1}).
\begin{itemize}
    \item \verb|e11()|, \verb|e22()|, \verb|e33()|: the diagonal weight generators.
    \item \verb|e12()|: the $E_{12}$ element.
    \item \verb|e21()|: the $E_{21}$ element.
    \item \verb|e23_12()|, \verb|e23_22()|: the $E_{23}$ element, per target branch.
    \item \verb|e32_12()|, \verb|e32_22()|: the $E_{32}$ element, per target branch.
    \item \verb|e13_12()|, \verb|e13_22()|: the $E_{13}$ element, per target branch.
    \item \verb|e31_12()|, \verb|e31_22()|: the $E_{31}$ element, per target branch.
\end{itemize}
The raising-action predicates \verb|E12|, \verb|E23_12|, \verb|E23_22|,
\verb|E13_12|, \verb|E13_22| (in \verb|U3basis.h|) identify the target state of
each generator.

\subsection{Physical states: \texttt{u3::physicalvector}}
(\verb|U3basis.h|.) The $\mathrm{SU(3)}\supset\mathrm{SO(3)}$ reduced state
$\ket{(\lambda\mu)\,\kappa\,L}$.

\textbf{Constructor:}
\begin{itemize}
    \item \verb|u3::physicalvector(irrep, L)|: a physical state of angular momentum $L$.
\end{itemize}

\textbf{Accessors:}
\begin{itemize}
    \item \verb|irr()|: the underlying irrep.
    \item \verb|lam()| / \verb|lambda()|: the SU(3) label $\lambda$.
    \item \verb|mu()|: the SU(3) label $\mu$.
    \item \verb|n13()|, \verb|n23()|, \verb|n33()|: the U(3) labels.
    \item \verb|L()|: the angular momentum.
    \item \verb|kappa()|, \verb|kappamax()|: the inner multiplicity of $L$.
\end{itemize}

\subsection{Free functions}
\textbf{Multiplicities:}
\begin{itemize}
    \item \verb|u3::Kappa(irrep,L)|: inner multiplicity $\kappa_{\max}$ (also \verb|kmax|, \verb|mult|).
    \item \verb|u3::Rho(A,B,C)|: outer multiplicity $\rho_{\max}$ of irrep $C$ in $A\otimes B$ product (\verb|U3decomposition.h|).
\end{itemize}

\textbf{Bases:}
\begin{itemize}
    \item \verb|u3::gelfandBasis(irrep)|: all Gelfand states of the irrep.
    \item \verb|u3::physicalBasis(irrep)|: all physical states with nonzero $\kappa$.
    \item \verb|u3::findGelfand(g,gArr)|: the index of a state within the \verb|gArr| Gelfand basis.
\end{itemize}

\textbf{Decomposition:}
\begin{itemize}
    \item \verb|u3::decomposition(A,B)|: the full Kronecker series, returning (list of $C$, list of $\rho_{\max}$) with $\sum_C\rho_{\max}(C)\dim C=\dim A\dim B$.
\end{itemize}

\textbf{Physical generator reduced matrix elements:} Each takes two
\verb|physicalvector| states (bra, ket) and two \verb|std::vector<int>|
out-parameters (which receive the zero-based multiplicity labels), and returns an
\verb|Eigen::MatrixXd| of shape $\kappa'_{\max}\times\kappa_{\max}$.
\begin{itemize}
    \item \verb|RME_L|: the angular-momentum RME, $\langle(\lambda\mu)\kappa'L'\|\mathbf{L}\|(\lambda\mu)\kappa L\rangle=\delta_{L'L}\sqrt{L(L{+}1)(2L{+}1)}$.
    \item \verb|RME_Q|: the quadrupole RME, carrying an extra $\sqrt{2L'+1}$ (Sec.~\ref{sub:conv}).
\end{itemize}

\textbf{Transformation:} Both expand a physical state \verb|p|
in the Gelfand basis \verb|gArr|, returning the canonical-state indices and an amplitude matrix with one orthonormal
column per $\kappa$.
\begin{itemize}
    \item \verb|u3::phys_to_cano(p,M,gArr)|: the expansion at a single projection $M$.
    \item \verb|u3::phys_to_cano_allM(p,gArr)|: the expansion over all $M$.
\end{itemize}

\subsection{Coefficient containers}
(\verb|Definitions.h|.) Every container provides
\verb|operator()(...)|, \verb|operator[](i)|, a member \verb|print()| method, and stream insertion via \verb|operator<<|. The indexing operators return a writable reference for a non-const C++
container and a value for a const container. They access the same
coefficients using different indexing schemes:
\begin{itemize}
    \item \verb|block(i0,...,id)| selects a coefficient by its multiplicity
    indices, in the order listed for each container below. Each index starts at
    zero and is smaller than its corresponding \verb|*max()| count.
    \item \verb|block[i]| selects a coefficient by a single flat index,
    $0\leq i<\verb|block.size()|$. For axis sizes $d_0,\ldots,d_{m-1}$,
    the equivalent flat index is
    $
      i=\sum_{j=0}^{m-1} i_j\prod_{k=j+1}^{m-1}d_k,
    $
    where an empty product is one. Thus, for a four-axis container,
    $i=((i_0d_1+i_1)d_2+i_2)d_3+i_3$.
    \item \verb|data()| exposes the underlying flat buffer;
    \verb|size()| gives its number of coefficients; and \verb|empty()|
    tests whether it contains no coefficients.
    \item \verb|print(s)| and \verb|s << block|: output a formatted table of all multiplicity indices and their corresponding values to the stream \verb|s| (defaulting to \verb|std::cout| for \verb|print()|).
\end{itemize}

In the equations below, $A,B,C,D$ denote $\rm U(3)$ irreps, and $AB,BC,AC,BD,CD$
denote selected intermediate irreps in the indicated tensor products, rather
than the tensor products themselves. The total irrep is denoted by $T$ for
three- and four-irrep recoupling. Each $\rho$
labels a particular occurrence of an output irrep in its coupling channel;
$\kappa$ resolves repeated angular momenta within an SU(3) irrep. The
multiplicity indices in these equations use the library's zero-based indexing.

\textbf{\texttt{U3U2}}: canonical coupling of $A\otimes B\to C$, reduced with
respect to U(2):
$
  \left\langle
  A a; B b || C c
  \right\rangle_{\rho},
$
where $a,b,c$ are their fixed U(2) sublabels
$[n_{12},n_{22}]$. Only the outer multiplicity varies within the container,
so \verb|block(rho)| and \verb|block[rho]| are identical.
\begin{itemize}
    \item \verb|rhomax()|: the outer multiplicity $\rho_{\max}$.
    \item \verb|operator()(rho)|: indexed by $\rho$.
    \item \verb|print(s = std::cout)|: prints the coefficient row over $\rho$.
\end{itemize}

\textbf{\texttt{U3SO3}}: physical coupling of $A\otimes B\to C$, reduced with
respect to SO(3):
$
  \left\langle A\,\kappa_A L_A;\,B\,\kappa_B L_B
  \middle\| C\,\kappa_C L_C\right\rangle_{\rho},
$
where $L_A,L_B,L_C$ are fixed angular momenta. The three inner multiplicities and the outer
multiplicity are the four container axes.
\begin{itemize}
    \item \verb|kappaAmax()|: inner multiplicity $\kappa_{A,\max}$.
    \item \verb|kappaBmax()|: inner multiplicity $\kappa_{B,\max}$.
    \item \verb|kappaCmax()|: inner multiplicity $\kappa_{C,\max}$.
    \item \verb|rhomax()|: the outer multiplicity $\rho_{\max}$.
    \item \verb|operator()(kappaA,kappaB,kappaC,rho)|.
    \item \verb|print(s = std::cout)|: tabular print over $(\kappa_A,\kappa_B,\kappa_C,\rho)$.
\end{itemize}

\textbf{\texttt{U3U6}}: $U$-recoupling:
$
 U\!\left(A,B,T,C;\,AB,\rho_{AB},\rho_{AB\_C};\,
 BC,\rho_{BC},\rho_{A\_BC}\right).
$
The labels $\rho_{AB\_C}$ and $\rho_{A\_BC}$ resolve the final couplings
$AB\otimes C\to T$ and $A\otimes BC\to T$, respectively.
\begin{itemize}
    \item \verb|rhoABmax()|: multiplicity of the $A\otimes B\to AB$ coupling.
    \item \verb|rhoAB_Cmax()|: multiplicity of the $AB\otimes C$ coupling.
    \item \verb|rhoBCmax()|: multiplicity of the $B\otimes C\to BC$ coupling.
    \item \verb|rhoA_BCmax()|: multiplicity of the $A\otimes BC$ coupling.
    \item \verb|operator()(rhoAB,rhoAB_C,rhoBC,rhoA_BC)|.
    \item \verb|print(s = std::cout)|: tabular print over $(\rho_{AB},\rho_{AB\_C},\rho_{BC},\rho_{A\_BC})$.
\end{itemize}

\textbf{\texttt{U3Z6}}: $Z$-recoupling:
$
 Z\!\left(B,A,T,C;\,AB,\rho_{AB},\rho_{AB\_C};\,
 AC,\rho_{AC},\rho_{AC\_B}\right).
$
The labels $\rho_{AB\_C}$ and $\rho_{AC\_B}$ resolve the couplings $AB\otimes C\to T$
and $AC\otimes B\to T$, respectively. 
\begin{itemize}
    \item \verb|rhoABmax()|: multiplicity of the $A\otimes B\to AB$ coupling.
    \item \verb|rhoAB_Cmax()|: multiplicity of the $AB\otimes C$ coupling.
    \item \verb|rhoACmax()|: multiplicity of the $A\otimes C\to AC$ channel.
    \item \verb|rhoAC_Bmax()|: multiplicity of the $AC\otimes B$ channel.
    \item \verb|operator()(rhoAB,rhoAB_C,rhoAC,rhoAC_B)|.
    \item \verb|print(s = std::cout)|: tabular print over $(\rho_{AB},\rho_{AB\_C},\rho_{AC},\rho_{AC\_B})$.
\end{itemize}

\textbf{\texttt{U3U9}}: 9-$\rm U(3)$ recoupling:
\begin{equation}
 \left[\begin{array}{cccc}
 A & B & AB & \rho_{AB} \\
 C & D & CD & \rho_{CD} \\
 AC & BD & T & \rho_{AC\_BD} \\
 \rho_{AC} & \rho_{BD} & \rho_{AB\_CD} &
 \end{array}\right].
\end{equation}
The rightmost entries label the row couplings, and the bottom entries label
the column couplings. In particular, $\rho_{AB\_CD}$ labels
$AB\otimes CD\to T$ and is named \verb|rhoABCD| in the code;
$\rho_{AC\_BD}$ labels $AC\otimes BD\to T$ and is named
\verb|rhoACBD|. The container's axis order is 
$
 (\rho_{AC},\rho_{BD},\rho_{AB\_CD},
   \rho_{AB},\rho_{CD},\rho_{AC\_BD})
$.
\begin{itemize}
    \item \verb|rhoACmax()|: multiplicity of the $A\otimes C\to AC$ coupling.
    \item \verb|rhoBDmax()|: multiplicity of the $B\otimes D\to BD$ coupling.
    \item \verb|rhoABCDmax()|: multiplicity of the $AB \otimes CD \to T$ route.
    \item \verb|rhoABmax()|: multiplicity of the $A\otimes B\to AB$ coupling.
    \item \verb|rhoCDmax()|: multiplicity of the $C\otimes D\to CD$ coupling.
    \item \verb|rhoACBDmax()|: multiplicity of the $AC \otimes BD \to T$ route.
    \item \verb|operator()(rhoAC,rhoBD,rhoABCD,rhoAB,rhoCD,rhoACBD)|.
    \item \verb|print(s = std::cout)|: tabular print over all six recoupling multiplicity axes.
\end{itemize}

In C++ the \verb|U3U9| result is a view into a thread-local scratch buffer that
the next database call overwrites; copy it before the next call. The other four
own their buffers.

\subsection{Coefficient databases}
All coupling and recoupling coefficients are computed on demand and cached inside singleton database instances:
\verb|u3::wigU3U2db()|, \verb|u3::wigU3SO3db()|, \verb|u3::u3u6db()|, \verb|u3::u3z6db()|, and \verb|u3::u3u9db()|.
The databases provide two ways to retrieve coefficients:
\begin{itemize}
    \item \textbf{Structured container getters} (\verb|get_U3U2|, \verb|get_U3SO3|, \verb|get_U3U6|, \verb|get_U3Z6|, \verb|get_U3U9|): return the dedicated multidimensional containers described above (\verb|U3U2|, \verb|U3SO3|, \verb|U3U6|, \verb|U3Z6|, \verb|U3U9|). These containers support multi-index coordinate access via \verb|operator()|, linear flat indexing via \verb|operator[]|, shape accessors (\verb|*max()|), and formatted stream printing.
    \item \textbf{Direct and flat-buffer getters} (\verb|get|): return contiguous buffers without container wrapper overhead.
\end{itemize}

The available getters across the five databases are:
\begin{itemize}
    \item \textbf{Canonical coupling database} (\verb|u3::wigU3U2db()|):
    \begin{itemize}
        \item \verb|get_U3U2(| $n_{13}^a,n_{23}^a,n_{33}^a,n_{12}^a,n_{22}^a,\ \dots^b,\ \dots^c$ \verb|)| $\to$ \verb|U3U2|: structured container over $\rho$.
        \item \verb|get(| $n_{13}^a,n_{23}^a,n_{33}^a,n_{12}^a,n_{22}^a,\ \dots^b,\ \dots^c$ \verb|)| $\to$ \verb|Eigen::RowVectorXd|: direct row vector of reduced coupling coefficients $\langle A a; B b || C c \rangle_\rho$ across all outer multiplicities $\rho = 0,\dots,\rho_{\max}-1$. Returns an empty vector if the coupling is forbidden or outside branching rules.
    \end{itemize}

    \item \textbf{Physical coupling database} (\verb|u3::wigU3SO3db()|):
    \begin{itemize}
        \item \verb|get_U3SO3(| $\lambda_1,\mu_1,L_1,\ \lambda_2,\mu_2,L_2,\ \lambda_3,\mu_3,L_3$ \verb|)| $\to$ \verb|U3SO3|: structured 4D container indexed by $(\kappa_A,\kappa_B,\kappa_C,\rho)$.
        \item \verb|get(| $\lambda_1,\mu_1,L_1,\ \lambda_2,\mu_2,L_2,\ \lambda_3,\mu_3,L_3$ \verb|)| $\to$ \verb|std::vector<double>|: flat coefficient buffer laid out in row-major order of $(\kappa_A,\kappa_B,\kappa_C,\rho)$, with size $\kappa_{A,\max}\kappa_{B,\max}\kappa_{C,\max}\rho_{\max}$.
    \end{itemize}

    \item \textbf{$U$-recoupling database} (\verb|u3::u3u6db()|):
    \begin{itemize}
        \item \verb|get_U3U6(| $A,B,ABC,C,AB,\ \rho_{AB},\rho_{AB\_C},\ BC,\rho_{BC},\rho_{A\_BC}$ \verb|)| $\to$ \verb|U3U6|: structured container indexed by $(\rho_{AB},\rho_{AB\_C},\rho_{BC},\rho_{A\_BC})$.
        \item \verb|get(| $A,B,ABC,C,AB,\ \rho_{AB},\rho_{AB\_C},\ BC,\rho_{BC},\rho_{A\_BC}$ \verb|)| $\to$ \verb|std::vector<double>|: flat buffer of length $\rho_{AB}\rho_{AB\_C}\rho_{BC}\rho_{A\_BC}$.
        \item \verb|get(| $n_{13}^a,n_{23}^a,n_{33}^a,\ \dots,\ n_{13}^{bc},n_{23}^{bc},n_{33}^{bc}$ \verb|)| (18 integers) $\to$ \verb|std::vector<double>|: raw-integer overload taking the partition labels of $A,B,ABC,C,AB,BC$ directly, computing intermediate $\rho$ maxima internally.
    \end{itemize}

    \item \textbf{$Z$-recoupling database} (\verb|u3::u3z6db()|):
    \begin{itemize}
        \item \verb|get_U3Z6(| $B,A,ABC,C,AB,\ \rho_{AB},\rho_{AB\_C},\ AC,\rho_{AC},\rho_{AC\_B}$ \verb|)| $\to$ \verb|U3Z6|: structured container indexed by $(\rho_{AB},\rho_{AB\_C},\rho_{AC},\rho_{AC\_B})$ (note the leading $B,A$ exchanging the first pair).
        \item \verb|get(| $B,A,ABC,C,AB,\ \rho_{AB},\rho_{AB\_C},\ AC,\rho_{AC},\rho_{AC\_B}$ \verb|)| $\to$ \verb|std::vector<double>|: flat buffer of length $\rho_{AB}\rho_{AB\_C}\rho_{AC}\rho_{AC\_B}$.
        \item \verb|get(| $n_{13}^b,n_{23}^b,n_{33}^b,\ \dots,\ n_{13}^{ac},n_{23}^{ac},n_{33}^{ac}$ \verb|)| (18 integers) $\to$ \verb|std::vector<double>|: raw-integer overload taking the partition labels of $B,A,ABC,C,AB,AC$ directly.
    \end{itemize}

    \item \textbf{9-$\mathrm{U(3)}$ recoupling database} (\verb|u3::u3u9db()|):
    \begin{itemize}
        \item \verb|get_U3U9(| $A,B,C,D,AB,CD,AC,BD,ABCD,$ $\rho_{AC},\rho_{BD},\rho_{ABCD},\rho_{AB},\rho_{CD},\rho_{ACBD}$ \verb|)| $\to$ \verb|U3U9|: structured container view. Multiplicity parameters follow the container's axis order: column intermediate channels $(\rho_{AC},\rho_{BD},\rho_{ABCD})$ followed by row intermediate channels $(\rho_{AB},\rho_{CD},\rho_{ACBD})$.
        \item \verb|get(| $A,B,C,D,AB,CD,AC,BD,ABCD,$ $\rho_{AB},\rho_{CD},\rho_{ABCD},\rho_{AC},\rho_{BD},\rho_{ACBD}$ \verb|)| $\to$ \verb|std::span<const double>| (or \verb|std::vector<double>|): flat buffer of length $\rho_{AC}\rho_{BD}\rho_{ABCD}\rho_{AB}\rho_{CD}\rho_{ACBD}$. \textbf{Note the multiplicity ordering difference}: in \verb|get|, intermediate row channels $(\rho_{AB},\rho_{CD},\rho_{ABCD})$ precede column channels $(\rho_{AC},\rho_{BD},\rho_{ACBD})$, whereas in \verb|get_U3U9| the column channels appear first to match internal storage strides.
        \item \verb|get(| $n_{13}^a,n_{23}^a,n_{33}^a,\ \dots,\ n_{13}^{abcd},n_{23}^{abcd},n_{33}^{abcd}$ \verb|)| (27 integers) $\to$ \verb|std::vector<double>|: raw-integer overload taking the partition labels of all 9 participating irreps.
    \end{itemize}
\end{itemize}

The multiplicity maxima passed to recoupling routines must match the exact \verb|Rho| values for the corresponding channels; a mismatch triggers an internal assertion. A forbidden or empty coupling returns an empty container or buffer. You can query its shape, but must not index it.

Each database singleton provides a \verb|clear()| member function (and \verb|u3::clear()| clears all five) to remove cached coefficient blocks while retaining allocated thread workspaces.

\subsection{Conventions and normalization}
\label{sub:conv}
Phase and normalization conventions follow Ref.~\cite{Dang2024EPJP1}; documented
values and the generation procedure are in \verb|test/reference/|.
\begin{itemize}
    \item \textbf{Multiplicity indexing:} Counts come from \verb|Kappa|/\verb|Rho|; container accessors use zero-based indices.
    \item \textbf{Canonical-coupling phase:} For each $\rho$ block the phase and normalization are fixed by the highest-weight reduced coefficients---the orthonormal null space of the system imposed by $E_{23}$ and $E_{12}$ annihilating the highest-weight coupled state. For $\rho_{\max}>1$ (and self-conjugate irreps) the solver fixes \emph{a} valid orthonormal $\rho$ basis (Row Reduced Echelon Form + Gram-Schmidt + positive leading sign); any alternative must be applied consistently to every $\rho$-carrying coefficient. Conjugation of a Gelfand state satisfies $C\ket{g} = (-1)^{n_{23}-n_{12}-n_{22}-n_{11}}\ket{g.\mathrm{conjugate}()}$ (\verb|gelfand::conjugate_phase()|).
    \item \textbf{Physical-to-canonical transformation:} A physical SO(3) state $\ket{(\lambda\mu)\,\kappa\,L\,M}$ is expanded in canonical Gelfand states $\ket{g}$ with $M = e_{11} - e_{22} = 2n_{11} - (n_{12} + n_{22})$. The highest-weight state $M=L$ belongs to the null space of $L_+$; $L_-$ fixes the lower-$M$ phases within a $\kappa$ multiplet. Distinct conjugate irreps obey $C\ket{(\lambda\mu)\,\kappa\,L\,M} = (-1)^{\lambda+\mu-L+M}\ket{(\mu\lambda)\,\kappa\,L,-M}$ (for $\lambda>\mu$ the transformation is built from the conjugate irrep with the phase factor $(-1)^{n_{22}+n_{11}-L}$). For a self-conjugate irrep $(\lambda,\lambda)$ conjugation acts as an $M$-independent symmetric matrix $D$ ($D^2=\mathbb{1}$) that may mix $\kappa$.
    \item \textbf{Generator reduced matrix element factor:} \verb|RME_L|/\verb|RME_Q| carry an extra $\sqrt{2L'+1}$, so $\langle\kappa L\|\mathbf{Q}\|\kappa'L'\rangle=(-1)^{L-L'}\langle\kappa'L'\|\mathbf{Q}\|\kappa L\rangle$.
    \item \textbf{Input checks.} Invalid labels or out-of-range indices abort via assertion.
\end{itemize}

\section{Python bindings}
\label{sec:python}

The \verb|u3lib| module mirrors the C++ API. Build it with
\verb|make python|, using an interpreter whose ABI matches the extension. Irreps
(\verb|su3.irrep|, \verb|irrep|), \verb|gelfand|, \verb|physicalvector|, the free
functions \verb|Rho|, \verb|Kappa|, \verb|decomposition|, \verb|gelfandBasis|,
\verb|physicalBasis|, the physical-generator reduced matrix elements \verb|RME_L|/\verb|RME_Q|, the
transformation \verb|phys_to_cano|/\verb|phys_to_cano_allM|, and the five database
accessors are all exposed with the same names and semantics as
Sec.~\ref{sec:classes}. The structured containers provide
\verb|__call__| (multiplicity-indexed), flat \verb|data()| as a NumPy array, and
formatted printing via \verb|print()| and \verb|__str__|; database access before \verb|initialize()| or after
\verb|finalize()| raises \verb|RuntimeError| rather than aborting. The \verb|U3U9|
container, a scratch-buffer view in C++, is copied into owning storage
automatically in Python.
For worked examples of irreps, bases, coupling, and recoupling, see the Jupyter
notebook \verb|python/tutorial.ipynb|. Build and toolchain details are in the
top-level \verb|README.md|.

\section{AI-ready interface}
\label{sec:agents}

\subsection{Skill catalog and orchestration}

The \verb|agents/| directory contains skills that guide an AI assistant through
\texttt{u3lib} calculations. The catalog includes an index
(\verb|agents/AGENTS.md|), shared runtime instructions, and a skill for each
capability. Each skill specifies the inputs, library call, conventions, and a
numerical check. When an AI assistant with code-execution access enters the library
and opens the \texttt{README.md} file, it is told to skip the rest of the file and 
go directly to the \verb|agents/| directory and use those instructions
to run the configured local library. The library calculations do not require a
network service.

The \verb|install| skill (\verb|agents/install/SKILL.md|) covers setup on a
fresh checkout. The agent first runs \verb|make config| to resolve or download
dependencies, followed by \verb|make| and \verb|make test|. It can then run the
calculation skills. Table~\ref{tab:prompts} gives a sample request for each
skill, with expected results checked against the library.

The assistant translates a request into library calls: it selects a skill,
prepares inputs in the required conventions, runs the compiled C++ or Python
routine, and reports the results with their labels and numerical check.
The library computes the coefficients, Casimirs, and matrix elements; the
assistant does not derive or approximate them. If a request is ambiguous or
asks for an undefined quantity, the assistant asks for clarification or
explains the problem. For example, a first-order Casimir is not defined for
an SU(3) irrep.

\begin{table*}[tp]
    \centering
    \small
    \begin{tabular}{p{3.1cm}p{13.2cm}}
    \hline\hline
    Skill & Example prompt (expected result) \\
    \hline
    \verb|dim| & ``Dimension of the SU(3) irrep $(2,1)$ -- equivalently the U(3) irrep $[3,1,0]$ or any shift such as $[4,2,1]$?'' ($15$; shift-invariant) \\
    \verb|casimirs| & ``$C_2,C_3$ for the SU(3) irrep $(3,1)$, and $C_1,C_2,C_3$ for the U(3) irrep $[4,2,1]$.'' (SU(3): $50/3,160/9$; U(3): $C_1{=}7,\,32/3,\,56/9$, U(3)-only) \\
    \verb|cano-list| & ``Enumerate the canonical Gelfand basis of $(2,0)$ with each state's $\epsilon,2\Lambda,2M$ labels.'' ($6$ states) \\
    \verb|phys-list| & ``List the physical SO(3) basis $(\kappa,L)$ of $(4,2)$ and verify the dimension sum rule.'' ($L{=}6,5,4,3,2,0$ with $\kappa{=}1,1,2,1,2,1$; $\sum\kappa(2L{+}1){=}60$) \\
    \verb|cano-generators| & ``Act with $E_{12}$ on the Gelfand state $\ket{2,0,0;2,0;1}$ of $(2,0)$; give the resulting state and matrix element.'' ($\ket{2,0,0;2,0;2}$ with $\sqrt2$) \\
    \verb|phys-generators| & ``Compute $\langle(2,0)\,\kappa{=}1\,L{=}2\|\mathbf{Q}\|(2,0)\,\kappa{=}1\,L{=}2\rangle$ and $\langle\|\mathbf{L}\|\rangle$.'' ($\sqrt{35/3}\approx3.416$ and $\sqrt{30}\approx5.477$) \\
    \verb|phys-to-cano| & ``Express $\ket{(2,0)\,\kappa{=}1\,L{=}2\,M{=}0}$ in the canonical Gelfand basis.'' (amplitudes $\sqrt{1/3},\sqrt{2/3}$) \\
    \verb|inner-mult| & ``SO(3) multiplicity $\kappa$ of $L{=}2$ in $(4,2)$?'' ($2$) \\
    \verb|outer-mult| & ``Outer multiplicity $\rho$ of $(1,1)$ in $(1,1)\otimes(1,1)$?'' ($2$) \\
    \verb|decomposition| & ``Full decomposition of $(2,0)\otimes(0,2)$ with outer multiplicities; check the dimension sum rule.'' ($(2,2){+}(1,1){+}(0,0)$) \\
    \verb|cano-coupling| & ``$\mathrm{U(3)}{\supset}\mathrm{U(2)}$ reduced coupling for highest-weight states $[2,0,0]\,[2,0]\times[2,0,0]\,[2,0]\to[4,0,0]\,[4,0]$ of $(2,0)\times(2,0)\to(4,0)$; verify it equals $1$.'' \\
    \verb|phys-coupling| & ``SO(3)-reduced coupling $(2,0)\,\kappa{=}1\,L{=}2\times(2,0)\,\kappa{=}1\,L{=}2\to(4,0)\,\kappa{=}1\,L{=}2$; check against $-\sqrt{2/9}$.'' \\
    \verb|u6recoupling| & ``$U$-recoupling for $(1,0)\otimes(1,0)\otimes(1,0)\to(3,0)$ through intermediate irreps $AB=(2,0)$ and $BC=(2,0)$.'' ($1$) \\
    \verb|z6recoupling| & ``$Z$-recoupling between $(AB)C$ and $(AC)B$ routes for $(1,0)\otimes(1,0)\otimes(1,0)\to(3,0)$ with intermediate irreps $AB=(2,0)$ and $AC=(2,0)$.'' ($1$) \\
    \verb|u9recoupling| & ``$9$-$(\lambda\mu)$ recoupling for four fundamentals $(1,0)$ coupling to $(4,0)$ with intermediate irreps $AB=CD=AC=BD=(2,0)$.'' ($1$) \\
    \verb|conjugation| & ``Conjugate the SU(3) irrep $(2,1)$ and its Gelfand states, reporting the phases.'' (irrep $\to(1,2)$, phase power $-1$) \\
    \hline\hline
    \end{tabular}
    \caption{The \texttt{u3lib} agent skills with a representative prompt for each.}
    \label{tab:prompts}
\end{table*}

\subsection{Evaluation objectives}
\label{app:benchmarking}
To evaluate the practical value of the skill catalog, we compare autonomous
coding agents with and without access to the skills. We ask whether the skills help agents complete tasks
correctly with fewer resources, and whether the benefit depends on the
implementation complexity of the task. We examine two hypotheses:
\begin{enumerate}
    \item {Resource efficiency:} Domain skills reduce token consumption,
    execution time, and API cost by supplying conventions and usage patterns
    that the agent would otherwise need to discover from the source.
    \item {Dependence on task complexity:} The benefit increases as tasks
    move from Python integration to C++ workspace management and parallel
    execution, where memory ownership and thread-safety requirements introduce
    additional constraints.
\end{enumerate}

\subsection{Experimental methodology}

\subsubsection{Benchmark Tasks}
We designed three tasks with increasing implementation complexity, covering Python integration, C++ workspace management, and parallel execution:
\begin{itemize}
    \item {Task 1: Python Racah recoupling (\texttt{consumer/recouple.py}).}
    A standalone Python utility computing $U$- and $Z$-recoupling coefficients through the pybind11 interface and verifying unitary symmetry relations across fundamental representations (e.g.\ $(1,0){\times}(1,0){\times}(1,0)\to(3,0)$). \emph{Architectural complexity: low} (declarative bindings, automatic memory, single-threaded).
    
    \item {Task 2: C++ 9-$\mathrm{U}(3)$ batch exporter (\texttt{consumer/u9\_batch.cpp}).}
    A C++ executable evaluating 9-$\mathrm{U}(3)$ blocks via \verb|u3::u3u9db()|: it parses the multi-irrep label specification, computes intermediate outer multiplicities, and serializes block and entry records, copying scratch-buffer views into owning storage to avoid aliasing across iterations. \emph{Architectural complexity: medium} (C++ typing, headers, manual workspace lifecycle).

    \item {Task 3: C++ multi-threaded coupling (\texttt{consumer/coupling\_batch.cpp}).}
    Parallel canonical- and physical-basis coupling with OpenMP (\verb|#pragma omp parallel for|), requiring per-thread database cache and scratch state (initialized on each thread) to avoid races, correct \verb|-fopenmp| configuration, and deterministic output. \emph{Architectural complexity: high} (shared-memory concurrency, thread-private state isolation).
\end{itemize}

\subsubsection{Isolated test environments}
Each run used a separate, disposable sandbox directory with its own consumer
build environment, so files from one trial were unavailable to subsequent
trials. The two conditions differed only in access to the skills:
\begin{itemize}
    \item {Control (zero skills):} a clean checkout containing the source, headers, and build scripts, with the skill directory (\verb|agents/|) removed.
    \item {Treatment (domain skills):} the same workspace plus the repository-local skill catalog indexed by \verb|agents/AGENTS.md|, which the prompt directed the agent to consult.
\end{itemize}

\subsubsection{Hidden tests and grader validation}
The agent could inspect files and directories, edit code, and execute shell
commands. When it declared completion or reached the limit of 50 steps, the
environment was locked and an automated grader tested the resulting program
against a hidden validation suite. The tests compared coupling and recoupling
values with exact group-theoretical results across a range of irreps, using a
$10^{-12}$ tolerance.

Before the trials, we tested the grader by introducing phase inversions,
memory leaks, uninitialized buffer reads, and artificial race conditions into
reference implementations. On these validation cases, the grader achieved
100\% sensitivity and specificity, with no false positives or false negatives.

\subsubsection{Model and Statistical Protocol}
All experiments used Anthropic's Claude Sonnet 5.5 (\texttt{claude-sonnet-5-5}) in an autonomous agent
test environment. For each task, $N = 10$ matched Control/Treatment pairs were run (60
runs total), recording four metrics per run: functional pass rate; total
token consumption (prompt, cache-read, and completion tokens); wall-clock
latency; and API cost at official token pricing. Significance was assessed with
two-tailed Welch's $t$-tests (unequal variance), reporting $t$, degrees of
freedom, and $p$.

\subsection{Results and discussion}

\subsubsection{Overall results}
\begin{figure*}[tp] 
    \centering
    \includegraphics[width=0.98\linewidth]{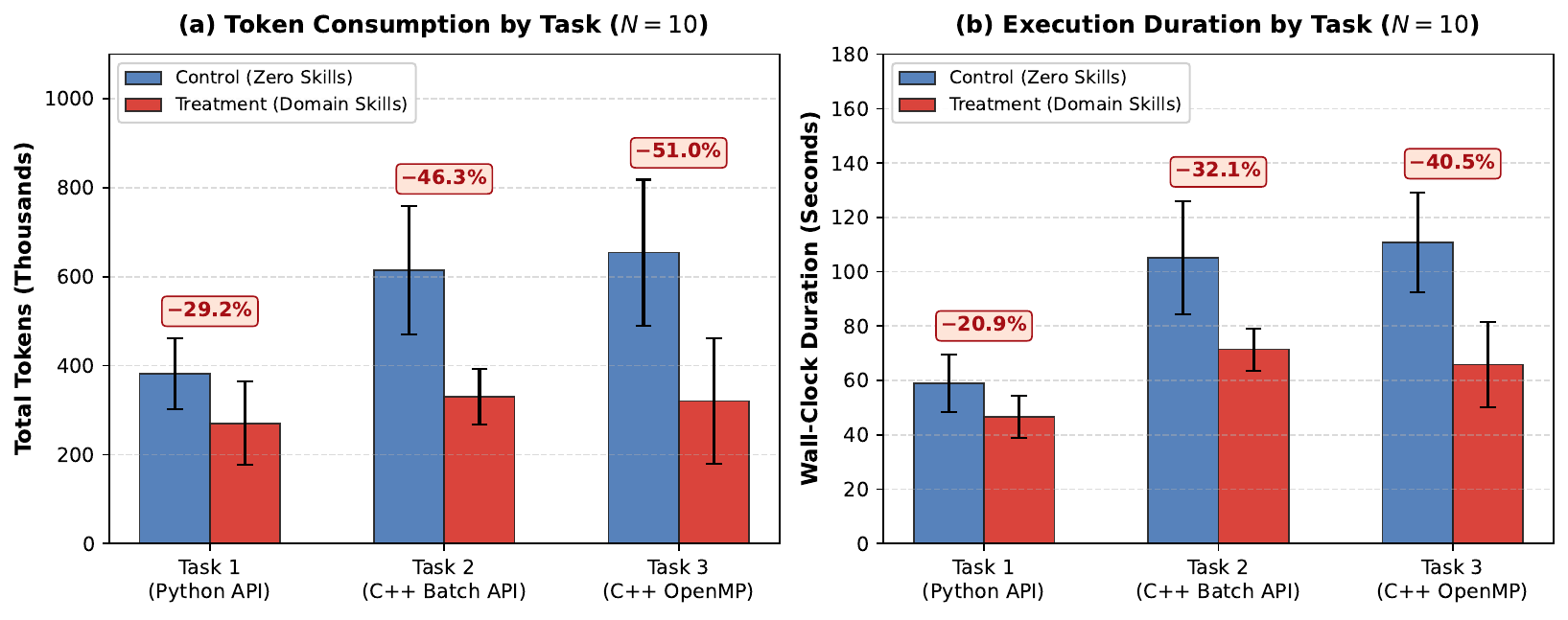}
    \caption{Benchmark results comparing autonomous agent performance between Control (Zero Skills) and Treatment (Domain Skills) across $N=10$ independent trials per task. (a) Total token consumption (thousands) by task, showing monotonic savings of 29.2\%, 46.3\%, and 51.0\%. (b) Wall-clock execution duration (seconds) by task, showing speedups of 20.9\%, 32.1\%, and 40.5\%. Error bars indicate $\pm 1$ sample standard deviation.}
    \label{fig:benchmark_summary}
\end{figure*}

Table~\ref{tab:benchmark_results} and Fig.~\ref{fig:benchmark_summary} summarize
the 60 runs. Across the three tasks:
\begin{itemize}
    \item {Token Consumption:} Providing domain skills reduced mean token consumption from $550,103 \pm 178,309$ tokens to $306,778 \pm 104,076$ tokens---an aggregate reduction of {44.2\%} ($t = -6.46$, $p = 6.0 \times 10^{-8}$, Welch's $t$-test).
    \item {Execution Latency:} Wall-clock execution time decreased from $91.61 \pm 28.85$\,s to $61.31 \pm 15.21$\,s, representing an aggregate speedup of {33.1\%} ($t = -5.09$, $p = 7.2 \times 10^{-6}$).
    \item {API Cost:} Direct inference costs fell from an average of \$0.3488 to \$0.2413 per task ({$-$30.8\%}, $t = -7.16$, $p = 4.4 \times 10^{-9}$).
    \item {Reliability:} Treatment runs achieved a 100\% pass rate across all 30 evaluations. One Control run failed Task 3 (9/10), giving an aggregate pass rate of 96.7\%.
\end{itemize}

\begin{table*}[tp]
    \centering
    \footnotesize
    \setlength{\tabcolsep}{2.8pt}
    \renewcommand{\arraystretch}{1.3}
    \caption{Empirical performance comparison between Control (Zero Skills) and Treatment (Domain Skills) across $N=10$ independent matched pairs per task (60 runs total) using Claude Sonnet 5.5. Tokens, execution duration, and costs are reported as $\text{Mean} \pm \text{Standard Deviation}$. Statistical significance is assessed via two-tailed Welch's $t$-tests.}
    \label{tab:benchmark_results}
    \vspace{0.2cm}
    \begin{tabular}{llcccccc}
    \hline\hline
    Task & Condition & Pass Rate & Total Tokens & Time (s) & Cost (\$) & Token $\Delta$ & Time $\Delta$ \\
    \hline
    \textbf{Task 1} 
      & Control ($N{=}10$)   & 100\% (10/10) & $381,613 \pm 80,140$ & $58.90 \pm 10.51$ & \$0.2780 & --- & --- \\
      & Treatment ($N{=}10$) & 100\% (10/10) & $270,260 \pm 93,890$ & $46.61 \pm 7.88$  & \$0.2280 & \textbf{$-$29.2\%}$^*$ & \textbf{$-$20.9\%}$^{**}$ \\
    \hline
    \textbf{Task 2} 
      & Control ($N{=}10$)   & 100\% (10/10) & $614,511 \pm 144,848$ & $105.21 \pm 20.86$ & \$0.3751 & --- & --- \\
      & Treatment ($N{=}10$) & 100\% (10/10) & $329,829 \pm 62,106$  & $71.42 \pm 7.74$   & \$0.2499 & \textbf{$-$46.3\%}$^{***}$ & \textbf{$-$32.1\%}$^{***}$ \\
    \hline
    \textbf{Task 3} 
      & Control ($N{=}10$)   & 90.0\% (9/10)  & $654,185 \pm 163,874$ & $110.73 \pm 18.35$ & \$0.3934 & --- & --- \\
      & Treatment ($N{=}10$) & 100\% (10/10) & $320,246 \pm 141,267$ & $65.89 \pm 15.71$  & \$0.2459 & \textbf{$-$51.0\%}$^{***}$ & \textbf{$-$40.5\%}$^{***}$ \\
    \hline\hline
    \textbf{Aggregate} 
      & Control ($N{=}30$)   & 96.7\% (29/30)  & $550,103 \pm 178,309$ & $91.61 \pm 28.85$ & \$0.3488 & --- & --- \\
      & Treatment ($N{=}30$) & 100\% (30/30) & $306,778 \pm 104,076$ & $61.31 \pm 15.21$ & \$0.2413 & \textbf{$-$44.2\%}$^{***}$ & \textbf{$-$33.1\%}$^{***}$ \\
    \hline\hline
    \multicolumn{8}{l}{\scriptsize $^*p < 0.05$ ($t = -2.85$); \quad $^{**}p < 0.01$ ($t = -2.96$); \quad $^{***}p < 0.001$ (Task 2: $t_{\text{tok}}{=}-5.71, t_{\text{time}}{=}-4.80$;} \\
    \multicolumn{8}{l}{\scriptsize Task 3: $t_{\text{tok}}{=}-4.88, t_{\text{time}}{=}-5.87$; \quad Grand Aggregate: $t_{\text{tok}}{=}-6.46, t_{\text{time}}{=}-5.09$; Welch's two-tailed $t$-test).}
    \end{tabular}
\end{table*}

\subsubsection{Dependence on task complexity}
For the three tasks considered here, the relative reductions in token consumption and execution time increase with the intended progression in task complexity:
\begin{align}
\Delta_{\text{tokens}} &= -29.2\% \text{ (Task 1)} \;\longrightarrow\; -46.3\% \text{ (Task 2)} \;\longrightarrow\; -51.0\% \text{ (Task 3)}, \notag \\
\Delta_{\text{time}}   &= -20.9\% \text{ (Task 1)} \;\longrightarrow\; -32.1\% \text{ (Task 2)} \;\longrightarrow\; -40.5\% \text{ (Task 3)}. \notag
\end{align}

This trend is consistent with the hypothesis that explicit guidance becomes more useful as implementation constraints increase. Three tasks alone do not establish a general scaling law; the following observations describe agent behavior within this benchmark:
\begin{enumerate}
    \item {Task 1 (API discovery):} Control agents produced correct recoupling solutions but typically needed 3--6 shell turns (\texttt{grep}, \texttt{ls}, inspecting pybind11 modules) to discover the exposed bindings (\verb|u3lib.u6recoupling|, \verb|u3lib.z6recoupling|) and their argument order. Treatment agents retrieved the signatures directly from the catalog, finishing in $16.6$ steps versus $22.8$ for Control.
    \item {Task 2 (scratch-buffer aliasing):} The 9-$\mathrm{U}(3)$ database (\verb|u3::u3u9db()|) returns views into internal scratch buffers reused across queries. Control agents encountered aliasing bugs when later queries overwrote stored blocks, spent time locating \verb|U3_U9Coeffs.h|, and misordered the multiplicity strides, inflating consumption to $614,511$ tokens on average (worst case exceeding $853,000$). Treatment agents used the documented copying and indexing procedures and completed the task in $329,829$ tokens with about half the variance.
    \item {Task 3 (concurrency):} The coefficient databases keep per-thread cache and scratch state, so sharing one database across threads without initializing each thread causes races. The second Control run failed the hidden tests after attempts to resolve non-deterministic segmentation faults; even successful Control runs averaged $33.6$ steps and $110.7$\,s to find the need for per-thread state. Treatment agents applied the skill's OpenMP guidance from the first turn, completing the task with deterministic output in $65.89$\,s and $320,246$ tokens.
\end{enumerate}

\section{Build and installation}
\label{sec:build}

\texttt{u3lib} is header-only; user programs include the headers in \verb|include/|
directly. The source is hosted publicly at
\url{https://github.com/Deep-SANE/u3lib}.
The build reads dependency paths from a configuration environment. After
sourcing it from the repository root, use the following make targets: \verb|make| (example programs $\to$ \verb|bin/|), \verb|make test|
(build and run the regression suite), and \verb|make python| (the pybind11
extension). The regression tests check dimensions, Gelfand interlacing and
ordering, SO(3) branching, canonical/physical coupling orthonormality across
$\kappa$ and $\rho$, conjugation identities, and documented analytic values;
comparisons use numerical tolerances, since results need not be bitwise
identical across backends.

\verb|make config| locates the dependencies and writes
\verb|config/config.generated.mk|. It uses existing installations or downloads
pinned versions from upstream sources and builds them under
\verb|external-libs/|, which is not part of the repository:
\begin{itemize}
    \item \texttt{libwignernj}: \url{https://github.com/susilehtola/libwignernj} 
    \item \texttt{Eigen3}: \url{https://eigen.tuxfamily.org}
    \item \texttt{Boost}: \url{https://www.boost.org}
    \item \texttt{OpenMP} (via the compiler runtime, e.g.\ GCC's libgomp): \url{https://www.openmp.org}
    \item \texttt{fmt}: \url{https://github.com/fmtlib/fmt}
    \item \texttt{pybind11}: \url{https://github.com/pybind/pybind11}
\end{itemize}

The \texttt{KraNS} null-space solver (\url{https://github.com/Deep-SANE/krans})
is used as a sibling checkout alongside \texttt{u3lib} rather than vendored under
\verb|external-libs/|; \verb|make config| clones it from the Deep-SANE
organization and builds it in place when it is not already present. 

\section{Conclusion}
\label{sec:conclusion}

In summary, we have presented \texttt{u3lib}, an open-source, header-only C++20 library with one-to-one Python bindings for $\rm U(3)$ symmetry algebra. Built on the generator-action and null-space procedure of Dang \textit{et al.}~\cite{Dang2024EPJP1} and paired with the iterative sparse solver \texttt{KraNS}~\cite{Krastenov2026MSc}, the framework evaluates canonical $\mathrm{U(3)}\supset\mathrm{U(2)}$ and physical $\mathrm{SU(3)}\supset\mathrm{SO(3)}$ coupling coefficients as well as 6- and 9-$\rm U(3)$ recoupling coefficients without the factorial-heavy alternating sums of traditional formulations. The library also provides irrep invariants, canonical Gelfand--Tsetlin and physical basis enumeration, generator matrix elements, basis transformations, and tensor-product decompositions. To support autonomous scientific workflows, the repository includes a structured catalog of domain skills that encode group-theoretical conventions, scratch-buffer lifetimes, and OpenMP thread-isolation rules for code-executing AI agents. Across 60 controlled benchmark runs spanning Python integration, C++ batch processing, and multi-threaded OpenMP execution, access to these skills raised task completion to 100\% while reducing mean token consumption by 44.2\%, wall-clock runtime by 33.1\%, and inference cost by 30.8\%, with the largest savings of 51.0\% in tokens and 40.5\% in runtime occurring on the concurrent C++ task. While three tasks on a single language model do not establish a universal scaling law and motivate broader evaluation across models and workloads, these results indicate that repository-local domain guidance becomes increasingly effective as implementation and concurrency constraints tighten. Within the broader \texttt{DeepSANE} ecosystem, \texttt{u3lib} provides the algebraic foundation for higher-rank extensions such as Wigner's $\rm U(4)$ supermultiplet library \texttt{u4lib} and symmetry-adapted \textit{ab initio} nuclear-structure framework \texttt{SAM}, while offering a reusable building block for $\rm U(3)$/$\rm SU(3)$ calculations in quantum information, condensed-matter physics, and lattice gauge theory.

\begin{acknowledgments}
This work is dedicated to the memory of J. P. Draayer, who passed on September 13, 2026 and whose pioneering contributions to algebraic computational nuclear physics through Elliott's $\rm SU(3)$ model, the Symplectic $\rm Sp(3,R)$ model, the \textit{ab initio} Symmetry-Adapted No-Core Shell Model, and his recent effort in Wigner's $\rm SU(4)$ model inspired this framework. The authors also gratefully acknowledge F. Pan and D. Kekejian for valuable discussions. PD was supported by the Institute of International Education through the Quad Fellowship, Louisiana State University, and Quantum CodeX. DL and TD were funded by Quantum CodeX. This work was supported in part by complimentary access to \href{https://antigravity.google/use-cases/science}{Google Antigravity} and Gemini 4 Argon to accelerate scientific workflows.
\end{acknowledgments}

\end{document}